\documentclass[a4paper,11pt]{amsart}
\begin{document}
\hyphenation{cha-rac-te-ris-tic ex-pe-ri-men-ta-lists sa-tis-fies
theo-ry ap-proxi-ma-tion}
\title[Non-existence of gravitational waves etc.]
{{\bf  Non-existence of gravitational waves\\ The stages of the theoretical discovery (1917-2003)}}
\author{Angelo Loinger}
\date{}
\address{Dipartimento di Fisica, Universit\`a di Milano, Via
Celoria, 16 - 20133 Milano (Italy)}
\email{angelo.loinger@mi.infn.it}

\begin{abstract}
A short history of the theoretical discovery that the
gravitational waves of general relativity do not have a physical
reality.
\end{abstract}

\maketitle

\vskip1.20cm
\noindent {\bf Introduction}\par \vskip0.10cm
I shall recall here the \emph{main} stages of an important
theoretical discovery: the general theory of relativity (GR) does
not allow the \emph{physical} existence of gravitational waves.
The solutions of the Einstein field equations which have a wave
character describe only formal undulations, quite destitute of a
physical reality.

\vskip0.80cm
\noindent {\bf 1917}
\par \vskip0.10cm
In this year Tullio Levi-Civita published a very fundamental
memoir ``On the analytic expression that must be given to the
gravitational tensor in Einstein's theory'' \cite{1}. The
conclusion of the paper is straightforward: the Einstein field
equations tell us that when the mass tensor $T_{jk}$ vanishes the
same occurrence must happen to the gravitational tensor $\left(1/
\kappa \right)\left(R_{jk}-\frac{1}{2}g_{jk}R\right)$. ``This fact
entails total lack of stresses, of energy flow, and also of a
simple localisation of energy''.
\par This result has an unquestionable \emph{\textbf{logical}}
soundness, as it was finally admitted by Einstein himself. Of
course, it implies the rejection of the various pseudo (false)
energy tensors of the gravitational field proposed by Einstein and
by other authors: a \emph{false} tensor cannot have a \emph{true}
physical meaning!
\par
Einstein objected that in such a way the total energy-momentum of
a closed system would always be equal to zero -- and this fact
would not imply the further existence of the system under whatever
form. However, from the standpoint of the \emph{coherence} of the
formalism, Levi-Civita  -- and Lorentz -- \cite{1} were
undoubtedly right. It is indeed sufficient to remember that in the
action principle of \emph{any} physical theory referred to
\emph{general} co-ordinates, the coefficients of the variations
$\delta h^{jk}$ of the metric tensor $h^{jk}$ are the components,
say $E_{jk}$, of the energy tensor of the considered field. But in
GR this property is just possessed, \emph{in vacuo}, by the tensor
$R_{jk}-\frac{1}{2}g_{jk}R$. \par The gravitational waves, as
objects without a \emph{true} energy-momentum, are only ghost
undulations.

\vskip0.80cm
\noindent {\bf 1930}
\par \vskip0.10cm
In this year Tullio Levi-Civita published an original study on the
characteristic hypersurfaces of Einstein field equations \cite{2}.
\par
He discovered that the functions $z(x)$,
$\left[x\equiv\left(x^{0},x^{1},x^{2},x^{3}\right)\right]$, of the
characteristic hypersurfaces $z(x)=0$ of Einstein field equations
are solutions of the Hamiltionian equation

\begin{equation} \label{eq:one}
    H:=\frac{1}{2}g^{jk}(x)\frac{\partial z(x)}{\partial x^{j}}\frac{\partial z(x)}{\partial x^{k}}=0
\end{equation}

According to Levi-Civita, the equation $z(x)=0$ gives the law of
motion of an \emph{electromagnetic} wave front -- or of the wave
front of \emph{any} field, capable of transmitting signals,
\emph{different} from the gravitational field. This interpretation
is quite obvious when $g^{jk}(x)$ has a \emph{non}-undulatory
form. If $g^{jk}(x)$ has a wavy form, there is no reason to
repudiate the above interpretation because \emph{the undulatory
character of} $g^{jk}(x)$ \emph{depends on the chosen system of
co-ordinates} \cite{3}. \par Remark that Levi-Civita's conception
is \emph{the} reasonable extension of that valid for the null
lines of \emph{special} relativity. Thus, also GR contains the
basic law of \emph{geometric optics} -- \emph{and independently of
Maxwell equations}.

\vskip0.80cm
\noindent {\bf 1953}
\par \vskip0.10cm
By means of perturbative computations, Scheidegger \cite{4} could
affirm that ``\ldots having explicitly shown that all the
radiation terms [of the gravitational field] whatsoever can be
destroyed by coordinate transformations, one observes that the
terms that have been found by straightforward calculations must be
entirely due to the particular choice of the coordinate system.
Thus there is no radiation damping of gravitational motion''. But
no damping means no emission of gravitational waves.

\vskip0.80cm
\noindent {\bf 1960}
\par \vskip0.10cm
In 1960 it was published an interesting book by Infeld and
Plebanski \cite{5}. At pages 200 and 201 we read: ``\ldots it is
hardly possible to connect any physical meaning with the flux of
energy and momentum tensor defined with the help of the
pseudo-energy-momentum tensor. Indeed, the [gravitational]
radiation can be annihilated by a proper choice of the coordinate
system. On the other hand, if we use a coordinate system in which
the flux of energy may exist, then it can be made whatsoever we
like by the addition of proper harmonic functions \ldots -- In the
linear theory we were faced with the choice between the retarded
and advanced potential. Here in the theory of gravitation the
choice is not so simple. Using the approximation procedure, we are
faced with the choice between single and double jumps. We can
speak only about [gravitational] radiation in the case of single
jumps. However, its existence or non-existence or its value will
depend upon the choice of arbitrary harmonic functions.'' We see
that also Infeld and Plebanski were quite sceptical about the real
existence of the gravitational waves.

\vskip0.80cm
\noindent {\bf 1998-2003}
\par \vskip0.10cm
In these years I have published several proofs of the
non-existence of the GW's; they are \emph{exact}, non-perturbative
proofs \cite{6}. I recall here only two demonstrations, which are
particularly simple.
\par \emph{i}) let us assume that at a given instant $t$ of its
motion a given point mass $M$ begins to send forth a GW, and let
us suppose that we know the \emph{kinematical characteristics} of
the motion between $t$ and $t+|\textrm{d}t|$. Then, we can
reproduce \emph{these} characteristics in a purely gravitational
motion of $M$ in a suitable ``external'', ``rigid'' gravitational
field, within a time interval equal to $|\textrm{d}t|$,
conveniently chosen. But in this case the mass $M$ moves along a
\emph{\textbf{geodesic}} -- and therefore it cannot emit any
gravitational radiation: indeed, the geodesic motions are ``free''
motions; they are the analogues of the rectilinear and uniform
motions of an electric charge of the customary Maxwell-Lorentz
theory. \par Thus we see that \emph{no} ``\emph{mechanism}''
\emph{exists for the generation of gravitational waves} -- the
above restriction to motions of mass \emph{points} is
\emph{conceptually} inessential. All the solutions of the
Einsteinian field equations having an undulatory character do
\emph{not} describe \emph{physical} waves \cite{7}.
\par \emph{ii}) As it is well known, in GR only the concepts and
the results that are \emph{independent} of the choice of the
system of \emph{\textbf{general}} co-ordinates have a
\emph{physical} meaning. Consider a solution of the Einstein field
equations which has -- in a given co-ordinate system -- a wavy
character. Through a \emph{finite} sequence of co-ordinate
transformations, endowed with convenient undulatory properties,
the primary undulating character of our solution can be completely
\emph{destroyed}. Thus, this character is only a property of the
original co-ordinate system, and therefore it has no physical
meaning. (According to a metropolitan legend, Bondi \emph{et al.}
-- cf., e.g., H. Bondi, F.A.E. Pirani and I. Robinson,
\emph{Proc.Roy.Soc.}, \textbf{A251} (1959) 219 -- would have
proved the existence of a class of privileged frames insofar as
the GW's are concerned. Now, their ``proof'' is fully destitute of
logical rigour, it is the mere expression of a desire. See at p.55
of my book quoted in \cite{6}).
\par I remark further that the propagation velocity of any metric
tensor depends on the reference system: with a suitable choice of
general co-ordinates, this velocity can take any value between
zero and infinite.

\normalsize \vskip0.5cm
\noindent {APPENDIX}\\ \emph{\textbf{On the linear approximation
of GR}}
\par \vskip0.15cm
If we restrict ourselves to the \emph{linear} approximation of GR
-- as the experimentalists generally do --, which has Minkowski
spacetime as its substrate, the physical existence of the GW's
seems, at first sight, a theoretical possibility. But the
energy-momentum of such GW's has a tensor character only under
Lorentz transformations, not  under \emph{general}
transformations. Therefore, it is always possible to find -- and
we remain, of course, in the ambit of the linearized version of GR
-- a general system of co-ordinates for which the above
energy-momentum is equal to \emph{zero}. \par In 1944 Weyl
published a remarkable article entitled ``How far can one get with
a linear field theory of gravitation in flat space-time?''
\cite{8}. He remarked, in particular, that Einstein's theory of
\emph{weak} gravitational fields (i.e., the linear approximation
of GR) resembles very closely Maxwell's theory of the e.m. fields,
and satisfies a principle of gauge invariance involving four
arbitrary functions, but its gravitational field exerts no force
on matter, i.e. it remains ``a powerless shadow''. From the
standpoint of the \emph{exact} GR, this is as it should be,
because ``the gravitational force arises only when one continues
the approximation beyond the linear stage''. Clearly, Weyl alludes
here a fundamental result of the EIH-method \cite{9}. Thus, we
find another argument -- and a strong argument -- against the
physical adequacy of the linearized version of GR insofar as the
question of the GW's is concerned.
\par I am very grateful to Prof. A. Gsponer, who has called my
attention to Weyl's paper.

\normalsize \vskip0.5cm
\noindent {PARERGON}\\ \emph{\textbf{On the PSR 1913+16}}
\par \vskip0.15cm
The overwhelming majority of the astrophysicists believe that the
time decrease of the revolution period of the binary radiopulsar
PSR 1913+16 gives an experimental (indirect) proof of the physical
reality of the gravitational radiation. As a matter of fact, the
perturbative quadrupole formula gives a decrease of the revolution
period which agrees very well with the observational data. \par I
emphasize the following points. \emph{i}) In the \emph{exact}
theory the quadrupole formula loses any meaning because the
hypothesized gravitational waves do \emph{not} have a \emph{true}
energy; therefore, the true mechanical energy which is lost during
the revolution motion ought to transform itself into the pseudo
(\emph{false}) energy of the hypothetical gravitational radiation:
the energy account does not balance. \emph{ii}) Many observational
astrophysicists know that realistic explanations of the decrease
of the revolution period are quite possible: for instance,
\emph{viscous losses} of the pulsar companion give a decrease of
the same order of magnitude of that given by the alleged emission
of gravitational waves. \emph{iii}) The empirical success of a
given theory -- or of a given computation -- is not an absolute
guaranty of its \emph{conceptual} adequacy: for instance, the
Ptolemaic theory of cycles and epicycles explained very well the
planetary orbits (with the only exception of Mercury's). \par The
serious scientists should abstain from wishful thinking.

\small \vskip0.5cm\par\hfill {\emph{``The king is naked!} -- cried
a child''.
     \vskip0.10cm\par\hfill (From an Andersen's tale)}

\end{document}